\documentclass{article}
\usepackage{epsfig}

\date{\today}

\usepackage{amsmath}
\usepackage{amssymb}

\begin{document}
\title{Stability of Gauss-Bonnet black holes in Anti-de-Sitter space-time
against scalar field condensation}
\author{{\large Yves Brihaye \footnote{email: yves.brihaye@umons.ac.be}
}$^{\ddagger}$ and 
{\large Betti Hartmann \footnote{email:
b.hartmann@jacobs-university.de}}$^{\dagger}$
\\ \\
$^{\ddagger}${\small Physique-Math\'ematique, Universite de
Mons-Hainaut, 7000 Mons, Belgium}\\ 
$^{\dagger}${\small School of Engineering and Science, Jacobs University Bremen,
28759 Bremen, Germany}  }

\date{}
\newcommand{\dd}{\mbox{d}}
\newcommand{\tr}{\mbox{tr}}
\newcommand{\la}{\lambda}
\newcommand{\ka}{\kappa}
\newcommand{\f}{\phi}
\newcommand{\vf}{\varphi}
\newcommand{\F}{\Phi}
\newcommand{\al}{\alpha}
\newcommand{\ga}{\gamma}
\newcommand{\de}{\delta}
\newcommand{\si}{\sigma}
\newcommand{\bomega}{\mbox{\boldmath $\omega$}}
\newcommand{\bsi}{\mbox{\boldmath $\sigma$}}
\newcommand{\bchi}{\mbox{\boldmath $\chi$}}
\newcommand{\bal}{\mbox{\boldmath $\alpha$}}
\newcommand{\bpsi}{\mbox{\boldmath $\psi$}}
\newcommand{\brho}{\mbox{\boldmath $\varrho$}}
\newcommand{\beps}{\mbox{\boldmath $\varepsilon$}}
\newcommand{\bxi}{\mbox{\boldmath $\xi$}}
\newcommand{\bbeta}{\mbox{\boldmath $\beta$}}
\newcommand{\ee}{\end{equation}}
\newcommand{\eea}{\end{eqnarray}}
\newcommand{\be}{\begin{equation}}
\newcommand{\bea}{\begin{eqnarray}}

\newcommand{\ii}{\mbox{i}}
\newcommand{\e}{\mbox{e}}
\newcommand{\pa}{\partial}
\newcommand{\Om}{\Omega}
\newcommand{\vep}{\varepsilon}
\newcommand{\bfph}{{\bf \phi}}
\newcommand{\lm}{\lambda}
\def\theequation{\arabic{equation}}
\renewcommand{\thefootnote}{\fnsymbol{footnote}}
\newcommand{\re}[1]{(\ref{#1})}
\newcommand{\R}{{\rm I \hspace{-0.52ex} R}}
\newcommand{\N}{{\sf N\hspace*{-1.0ex}\rule{0.15ex}%
{1.3ex}\hspace*{1.0ex}}}
\newcommand{\Q}{{\sf Q\hspace*{-1.1ex}\rule{0.15ex}%
{1.5ex}\hspace*{1.1ex}}}
\newcommand{\C}{{\sf C\hspace*{-0.9ex}\rule{0.15ex}%
{1.3ex}\hspace*{0.9ex}}}
\newcommand{\eins}{1\hspace{-0.56ex}{\rm I}}
\renewcommand{\thefootnote}{\arabic{footnote}}

\maketitle

\bigskip

\begin{abstract}
We study the stability of static, hyperbolic Gauss-Bonnet black holes in
$(4+1)$-dimensional Anti-de-Sitter (AdS) space-time
against the formation of scalar hair. Close to extremality the black holes
possess a near-horizon
topology of AdS$_2\times H^3$ such that within a certain range of the scalar
field mass one would expect that they become unstable
to the condensation of an uncharged scalar field. We confirm this numerically
and observe that there exists a family 
of hairy black hole solutions
labelled by the number of nodes of the scalar field function. We construct
explicit examples of solutions with a scalar field that
possesses zero nodes, one node and two nodes, respectively, and show that the
solutions
with nodes persist in the limit of Einstein gravity, i.e. for vanishing
Gauss-Bonnet coupling. We observe that
the interval of the mass for which scalar field condensation appears decreases
with increasing Gauss-Bonnet coupling
and/or with increasing node number.  
\end{abstract}
\medskip
\medskip
 \ \ \ PACS Numbers: 04.70.-s,  04.50.Gh, 11.25.Tq
\section{Introduction}
Higher curvature corrections appear naturally in the low energy effective action
of string theory \cite{zwiebach}. In more than
four dimensions the quadratic correction is often chosen to be the Gauss-Bonnet
(GB) term, which has the property
that the equations of motion are still second order in derivatives of the metric
functions. As such explicit
solutions of the equations of motion are known. The first example of static,
spherically symmetric and asymptotically flat
black hole solutions in GB gravity were given for the uncharged case in
\cite{deser,Wheeler:1985nh} and for the charged case in 
\cite{wiltshire}.
Moreover, the corresponding solutions in asymptotically Anti-de Sitter (AdS)
\cite{Cai:2001dz,Cvetic:2001bk,Cho:2002hq,Neupane:2002bf}
as well as de Sitter (dS) space-times \cite{Cai:2003gr} have been studied. In
most cases, black holes not only with
spherical ($k=1$), but also with flat ($k=0$) and hyperbolic ($k=-1$) horizon
topology have been considered.
Moreover, the thermodynamics of these black holes has been studied in detail
\cite{Cho:2002hq,Neupane:2002bf,Neupane:2003vz}
and the question of negative entropy for certain GB black holes in dS and AdS
has been discussed \cite{Cvetic:2001bk,Clunan:2004tb}.

The gravity--gauge theory duality \cite{ggdual} has attracted a lot of attention
in the past years. The most famous
example is the AdS/CFT correspondence \cite{adscft} which states that a gravity
theory in a $d$-dimensional
Anti-de Sitter (AdS) space--time is equivalent to a Conformal Field Theory (CFT)
on the $(d-1)$-dimensional boundary of AdS.
Recently, this correspondence has been used to describe so-called holographic
superconductors with the help of black holes in
higher dimensional AdS space--time \cite{gubser,hhh,horowitz_roberts,reviews}.
In most cases $(3+1)$-dimensional
black holes with planar horizons ($k=0$) were
chosen to account for the fact that high temperature superconductivity is mainly
associated to 2-dimensional layers within the material. 
The basic idea is that at low temperatures a planar black hole in asymptotically
AdS becomes unstable to the condensation
of a charged scalar field. The hairy black hole is the gravity dual of the
superconductor. The main point here is that
this instability occurs due to the fact that the scalar field is charged and its
effective mass
drops below the Breitenlohner-Freedman (BF) bound \cite{bf} for sufficiently low
temperature of the black hole hence
spontaneously breaking the U(1) gauge symmetry.
Surprisingly, however, the scalar condensation can also occur for uncharged
scalar fields in the $d$-dimensional planar Reissner-Nordstr\"om-AdS (RNAdS)
black hole 
space-time \cite{hhh}. This is a new type of instability that is not connected
to a spontaneous symmetry breaking as in the charged case.
Rather it is related to the fact that the planar RNAdS black hole possesses an
extremal limit
with vanishing Hawking temperature and near-horizon geometry 
AdS$_2\times {\mathbb{R}}^{d-2}$ (with $d \ge 4$)
\cite{Robinson:1959ev,Bertotti:1959pf,Bardeen:1999px}. For scalar field masses 
larger than the $d$-dimensional BF bound, but smaller than the $2$-dimensional
BF bound the near-horizon geometry
becomes unstable to the formation of scalar hair, while the asymptotic AdS$_d$
remains stable \cite{hhh}.
The fact that the near-horizon geometry of extremal black holes is a topological
product of two manifolds with constant
curvature has let to the development of the entropy function formalism
\cite{Sen:2005wa,sen2,dias_silva}. 
In \cite{Dias:2010ma} the question of the condensation of a uncharged scalar
field on uncharged black holes in $(4+1)$ dimensions has been addressed.
As a toy model for the rotating case, static black holes with hyperbolic
horizons ($k=-1$) were discussed.
In contrast to the uncharged, static black holes with flat ($k=0$) or spherical
($k=1$) horizon topology
hyperbolic black holes possess an extremal limit with near-horizon geometry
AdS$_2\times H^3$ and it has been shown
by numerical construction that the black holes form scalar hair close to
extremality.

Interestingly, there seems to be
a contradiction between the holographic superconductor approach and the
Coleman-Mermin-Wagner theorem \cite{CMW} 
which forbids spontaneous symmetry
breaking in $(2+1)$ dimensions at finite temperature. Consequently, it has been
suggested that 
higher curvature corrections and in particular GB terms should 
be included on the gravity side and holographic GB superconductors in $(3+1)$
dimensions have been
studied \cite{Gregory:2009fj}. However, though the critical temperature gets
lowered
when including GB terms, condensation cannot be suppressed -- not even when
including backreaction \cite{Brihaye:2010mr,Barclay:2010up}.

In this paper, we are interested in the condensation of an uncharged scalar
field on an uncharged hyperbolic GB 
black hole in $(4+1)$-dimensional AdS. These black holes have been shown to
possess a near-horizon geometry of AdS$_2\times H^3$ 
in the extremal limit \cite{aste}.  In accordance with the results in the
Einstein gravity limit \cite{Dias:2010ma} we would
thus expect that the hyperbolic GBAdS black holes would become unstable for
scalar field masses
below the $2$-dimensional BF bound. We construct explicit examples of hairy
GBAdS black holes and study their properties.
We also reinvestigate the Einstein gravity limited studied before  and show that
next to the 
solutions found in \cite{Dias:2010ma} their exists a family of hairy black hole
solutions labelled by the number
of nodes of the scalar field function. Our paper is organised as follows: in
Section 2 we give
our model, the equations of motions, the boundary conditions and discuss the
black hole properties
as well as exact solutions of the equations of motion. In Section 3 we give our
numerical
results, first for the linear case and then for the non-linear case. We conclude
in Section 4.

\section{The Model}
\label{model}
In this paper, we study Gauss-Bonnet (GB) black holes in $(4+1)$-dimensional
Anti-de Sitter 
(AdS) space-time against condensation of an uncharged scalar field. The action
that we consider reads~:
\begin{equation}
S= \frac{1}{16\pi G} \int d^5 x \sqrt{-g} \left(R -2\Lambda + 
\frac{\alpha}{2}\left(R^{MNKL} R_{MNKL} - 4 R^{MN}
R_{MN} + R^2\right) + 16\pi G {\cal L}_{\rm matter}\right) \ ,
\end{equation}
where $\Lambda=-6/L^2$ is the cosmological constant, $L$ denotes the AdS radius
and $G$ is Newton's constant.
$\alpha$ is the Gauss--Bonnet coupling that has dimension $({\rm length})^2$ and
is bounded from above $\alpha \le L^2/4$, where $\alpha=L^2/4$ is the
Chern-Simons
limit. $R$, $R_{MN}$ and $R_{MNKL}$, $M,N,K,L=0,1,2,3,4$ denote the Ricci scalar, the
Ricci tensor and the Riemann tensor, respectively, while
$g$ is the determinant of the metric. The matter Lagrangian ${\cal L}_{\rm
matter}$ of a real valued scalar field with mass $m^2$ reads~:
\begin{equation}
{\cal L}_{\rm matter}= - \left(\partial_M\psi\right) \partial^M \psi - m^2
\psi^2  \ \ , \ \  M=0,1,2,3,4  \ .
\end{equation}
In the following, we want to study static, hyperbolic black holes. 
The Ansatz for the metric hence reads~:
\begin{equation}
\label{ansatz}
ds^2 = - f(r) a^2(r) dt^2 + \frac{1}{f(r)} dr^2 + \frac{r^2}{L^2} d\Xi^2_{3}
\end{equation}
where $d\Xi^2_{3}$ is the line element of a 3-dimensional hyperbolic space with
curvature $k=-1$ and Ricci scalar $R=-6$.
The coupled Einstein and Euler--Lagrange equations are obtained from the
variation of the
action with respect to the matter and metric fields, respectively. They read~:
\begin{eqnarray}
\label{eq1}
     f' &=& \frac{2r(-1-f+ \frac{2 r^2}{{L^2}})}{r^2 - 2 \alpha(1+f)}  
     - \gamma r^3 
     \left(\frac{  m^2  \psi^2  +  f  \psi'^2}{r^2 - 2 \alpha (1+f)}\right) \ ,
\\ 
\label{eq2}
        a' &=& \gamma \frac{r^3 a  \psi'^2}{ r^2- 2 \alpha(1+f)} \ , \\
\label{eq3}
    \psi'' &=& -\left(\frac{3}{r} + \frac{f'}{f} + \frac{a'}{a}\right) \psi' +
\frac{m^2}{f} \psi \ ,
\end{eqnarray}
where we have used the abbreviation $\gamma=\frac{16}{3}\pi G$. Here and in the
following the prime denotes
the derivative with respect to $r$. Note that the equations
(\ref{eq1})-(\ref{eq3}) can be rescaled 
as follows $\psi \rightarrow \psi/\sqrt{\gamma}$ for $\gamma\neq 0$. We can
hence set $\gamma=1$ without loss of generality.
Different choices of $\gamma$ will simply lead to a different choice of
normalisation of $\psi$. 
For the linear case $\gamma=0$ the normalisation has to be fixed by hand.
  
The set of coupled, non-linear ordinary differential equations
(\ref{eq1})-(\ref{eq3}) has to be solved
numerically subject to appropriate boundary conditions. At the event horizon
$r=r_+$ these read
\begin{equation}
\label{bc1}
 f(r_+)=0  
\end{equation}
with $a(r_+)$ finite. Moreover, 
in order for the scalar field to be regular at the horizon we need to impose
\begin{equation}
\label{bc_horizon}
\psi'(r_+)=\left.\frac{m^2 \psi (r^2-2\alpha)}{-2r + 4r^3/L^2 - \gamma r^3 m^2
\psi^2}\right\vert_{r=r_+} \ .
\end{equation}
Asymptotically, we want the scalar field to have the following behaviour
\begin{equation} 
  \psi(r\gg 1) = \frac{\psi_+}{r^{\Delta_+}} + \frac{\psi_-}{r^{\Delta_-}} \ \
\label{decay}
\end{equation}
with
\begin{equation}
\label{lambda}
       \Delta_{\pm} = 2 \pm \sqrt{4+ m^2 L_{\rm eff}^2} \ \ , \ \ 
       L_{\rm eff}^2 \equiv \frac{2 \alpha}{1 - \sqrt{1 - 4 \alpha/L^2}}   \ .
\end{equation}
$L_{\rm eff}$ is the effective AdS radius, which behaves like $L_{\rm eff}^2\sim
L^2 \left(1  -  \alpha/ L^2 + O(\alpha^2)\right)$ for small
$\alpha$ and becomes equal to $L_{\rm eff}^2 = L^2/2$ in the Chern-Simons limit
$\alpha=L^2/4$. Hence, the effective
AdS radius decreases with increasing $\alpha$. Finally, the behaviour of the
metric functions
at $r\gg 1$ is given by
\begin{equation}
\label{expansionmetric}
 f(r\gg1)=-1 +\frac{r^2}{L_{\rm eff}^2} + \frac{f_2}{r^2} + O(r^{-4}) \ \ , \ \
a(r\gg 1)=1+\frac{a_4}{r^4} + O(r^{-6}) \ ,
\end{equation}
where $f_2$ and $a_4$ are constants that have to be determined numerically.

\subsection{Black hole properties}
The Hawking temperature $T_{\rm H}$ and entropy $S$ of the black holes studied
in this paper read \cite{aste}
\begin{equation}
\label{TS}
 T_{\rm H}=\frac{f'(r_+)a(r_+)}{4\pi} \ \ \ , \ \ \ \frac{S}{V_3}=
\frac{r_+^3}{4G}\left(1-\frac{6\alpha}{r_+^2}\right) \ \ \ ,  \ \ \ 
\end{equation}
where $V_3$ corresponds
to the volume of the 3-dimensional hyperbolic space with line element
$d\Xi_3^2$, which we assume to be compactified. In the following
we will set $V_3=1$ hence assuming that all quantities containing this factor
are given per unit
hyperboloid volume. The specific heat is 
given by $C=T_{\rm H}\left(\partial S/\partial T_{\rm H}\right)$. For $C > (<) \
0$ the black holes are thermodynamically stable (unstable).

We will also be interested in the energy $E$ of the solutions. Using the
counterterm approach by Brown and York 
\cite{brown,brihaye_radu} 
we find that the energy $E$ can be expressed in terms of the expansion
coefficients of the metric functions $f(r)$ and $a(r)$ 
(see (\ref{expansionmetric}))
as follows
\begin{equation}
\frac{16\pi G E}{V_3} = \sqrt{1-\frac{\alpha}{L^2}} \left(-3 f_2 - 8
\frac{a_4}{L_{\rm eff}^2}\right)  \ .
\end{equation}

The main feature that distinguishes Gauss-Bonnet-Anti-de Sitter (GBAdS) and
Schwarzschild-Anti-de Sitter (SAdS) black holes with hyperbolic
horizons from those with flat or spherical horizons is that an extremal limit
with $T_{\rm H}=0$ 
exists (see also \ref{exact}). 
Close to extremality the horizon topology is AdS$_{2}\times H^{3}$ with 
radius of the AdS$_2$ given by $R$ \cite{aste}. 
Now the Breitenlohner-Freedman bound \cite{bf} for a scalar field in
$d$-dimensional
AdS is
\begin{equation}
\label{BFd}
 m^2 \ge m_{\rm BF,d}^2 = - \frac{(d-1)^2}{4L_{\rm eff}^2}   \ .
\end{equation}
Using the entropy function formalism \cite{Sen:2005wa,sen2,dias_silva} we find
that the radius $R$ of the AdS$_2$ in this case reads (see
Appendix for details on the calculation for horizons with spherical, hyperbolic
and flat horizon topology)
\begin{equation}
 R= \sqrt{\frac{L^2}{4}-\alpha} \ .
\end{equation}
It is obvious from this formula that we need to require $L^2 > 4\alpha$ in order
to find an extremal solution
whose near-horizon topology possesses an AdS$_2$-factor. In the Chern-Simons
limit $L^2=4\alpha$, the radius of
the AdS$_2$ simply vanishes. Hence for scalar masses satisfying
\begin{equation}
\label{bf}
  - \frac{4}{L_{\rm eff}^2} \le m^2 \le -\frac{1}{4R^2}=-\frac{1}{L^2-4\alpha}
\end{equation}
we find that the asymptotic AdS$_5$ is stable, while the near horizon AdS$_2$ is
unstable.
We would thus expect that the GB black hole becomes unstable to the formation of
scalar hair in this
range of the parameter $m$.

\subsection{Exact solutions}
\label{exact}
In the case $\gamma=0$, i.e. when the scalar and gravity equations are
decoupled, there exists an explicit solution
to the equations \cite{deser,Wheeler:1985nh,Cai:2001dz}. The metric functions of
this Gauss-Bonnet-Anti-de Sitter (GBAdS) solutions read
\begin{equation}
\label{solu}
      f(r) = -1 + \frac{r^2}{2\alpha} \left(1-\sqrt{1-\frac{4\alpha}{L^2} 
+ \frac{4\alpha M}{r^4}} \right) = -1 + \frac{r^2}{2\alpha}
\left[ 1 -\sqrt{1+ \frac{4\alpha}{L^2}\left(\frac{r_+^4}{r^4}-1\right) +
\frac{4\alpha}{r^4}\left(\alpha-r_+^2\right)} \right] 
\end{equation}
and $a(r)\equiv 1$. $M$ is an integration constant that is connected to the
Arnowitt-Deser-Misner (ADM) mass of the solution $M_{\rm ADM}$ by
$M_{\rm ADM}=3MV_3/(16\pi G)$ \cite{Neupane:2003vz}. 
In the last equality in (\ref{solu}) we have used the relation between the event
horizon $r_+$ and $M$, where the horizons
of the space-time are given by
\begin{equation}
 r_{\pm} = \frac{1}{2}\sqrt{2L^2 \pm 2 \sqrt{L^4-4\alpha L^2 +4M L^2}}  \ 
\end{equation}
such that
\begin{equation}
\label{mass}
 M=\alpha + \frac{r_+^4}{L^2} - r_+^2  \ . 
\end{equation}

In the $\gamma=0$ limit, the scalar field equation (\ref{eq3}) becomes linear
and describes a scalar field
in the background space-time with metric functions given by (\ref{solu}) and
$a(r)\equiv 1$.
Note that $\psi(r)\equiv 0$ together with (\ref{solu}) is a solution to the full
set of equations
(\ref{eq1})-(\ref{eq3}) for generic $\gamma$. 
The Hawking temperature of this black hole reads
\begin{equation}
\label{temp}
 T_{\rm H} = \frac{1}{4 \pi} f'(r_+) a(r_+) = \frac{-r_+ +
\frac{2r_+^3}{L^2}}{2\pi(r_+^2 - 2\alpha)} \ .
\end{equation}
The extremal solution with $T_{\rm H}=0$ has horizon $r^{{\rm
(ex)}}_+=L/\sqrt{2}$ and mass parameter $M^{\rm ( ex)}=\alpha-L^2/4$, which
is always negative for $\alpha < L^2/4$. In order for the temperature $T_{\rm
H}$ to be non-negative, we have to
require that $r_+ \ge r_+^{\rm (ex)} \ge  \sqrt{2\alpha}$ which is always
fulfilled for $\alpha < L^2/4$.  
The energy of the GBAdS black hole solutions is given by
\begin{equation}
\label{energy}
 E=\sqrt{1-\frac{\alpha}{L^2}} M_{\rm ADM} \ .
\end{equation}
There are solutions with $E=M=M_{\rm ADM}=0$. These are vacuum solutions, which
have horizons at $r_+=r_{+,0}$ with
\begin{equation}
 \frac{r_{+,0}^2}{L^2} =\frac{1}{2}\left(1+\sqrt{1-\frac{\alpha}{4L^2}}\right) \
\end{equation}
such that $r_{+,0} > r_+^{\rm (ex)}$. Black holes with $r_+ < r_{+,0}$ have
negative energy and the extremal
solution has the lowest energy. 
The $M=0$ solution is special as it is locally isometric to AdS$_5$ with
effective AdS radius $L_{\rm eff}$. 
When the AdS$_5$ BF bound is satisfied (see (\ref{BFd})), i.e. when
$\Delta\equiv \Delta_{\pm}=2$ there is an
exact solution to the equations of motion with $\psi(r)=L_{\rm eff}^2/r^2$. Very
similar to the $\alpha=0$ case \cite{Dias:2010ma} this
does, however, not signal an instability of AdS$_5$ since our solution covers
only part of AdS$_5$. 

The entropy is given by the expression in (\ref{TS}). Note that this explicit
form can be derived from the first law
of black hole thermodynamics $dM_{\rm ADM}=T dS$ by using (\ref{mass}). Since
the entropy can become negative it has been
suggested in \cite{Clunan:2004tb} that the entropy $S$ of the vacuum solution
with $r_+=r_{+,0}$ should be
substracted from this expression assigning entropy $S=0$ to the vacuum solution
such that
\begin{equation}
 S=\int\limits_{r_{+,0}}^{r_+} T^{-1} \left(\frac{\partial M_{\rm ADM}}{\partial
r_+}\right) dr_+ \ .
\end{equation}

The specific heat $C$ can be calculated explicitly and reads
\begin{equation}
 4G V_3 C=\frac{3r_+ (2\alpha-r_+^2)^2 (2r_+^2-L^2)}{2r_+^4 +
r_+^2(L^2-12\alpha) + 2\alpha L^2} \ .
\end{equation}
This is always positive for $r_+ \ge r_+^{\rm (ex)}$ and $\alpha <L^2/4$. Hence,
the black holes
are thermodynamically stable.

There are two important limits of (\ref{solu}). The first is the Einstein
gravity limit $\alpha=0$, whose
stability with respect to the formation of scalar hair has been discussed in
detail in \cite{Dias:2010ma}. In this case the equations admit
Schwarzschild-Anti-de Sitter (SAdS) solutions with metric function $f(r)$ 
\begin{equation}
\label{SADS}
 f(r)=-1 + \frac{r^2}{L^2} - \frac{M}{r^2} = -1 + \frac{r^2}{L^2} +
\frac{r_+^2}{r^2}\left(1-\frac{r_+^2}{L^2}\right) \ . 
\end{equation}
This solution has an extremal limit with $r_+=r^{{\rm (ex)}}_+=L/\sqrt{2}$ and
mass parameter $M^{\rm ( ex)}=-L^2/4$.
The fact that the near-horizon topology of the SAdS solution is AdS$_2\times
H^3$ was used in \cite{Dias:2010ma} to
show that these black holes become unstable to the formation of scalar hair on
the horizon for specific values
of the mass parameter $m$.

The other limit is the Chern-Simons limit $\alpha=L^2/4$ with
\begin{equation}
\label{CS}
 f(r)=\frac{2}{L^2}\left(r^2-r_+^2\right) \ .
\end{equation}
The solution with $T_{\rm H}=0$ and $r^{{\rm (ex)}}_+=L/\sqrt{2}$ is special in
this case \cite{Neupane:2003vz}
since $M^{\rm (ex)}=0$. Along with \cite{Neupane:2003vz} we will hence restrict
ourselves to $\alpha < L^2/4$ in the following, also
because -- as mentioned above -- the AdS$_2$ in the Chern-Simons limit has
vanishing AdS radius $R$.

\section{Numerical results}
We have solved the set of equations (\ref{eq1})-(\ref{eq3}) subject to the
appropriate boundary conditions
numerically using a collocation method for
ordinary differential equations \cite{colsys}. The relative errors of the
solutions are on the order of $10^{-10}$ - $10^{-6}$.
In the following we will fix $r_+=1$ without loss of generality and restrict
ourselves to $\alpha  < 0.5$ to ensure that
the Hawking temperature $T_{\rm H}$ in the linear case $\gamma=0$ is positive
(see (\ref{temp})). We will also limit 
our analysis to $\Delta \ge 1$ to ensure that the modes are normalisable. 

Along with \cite{Dias:2010ma} we choose the scalar field to fall off
like $\psi(r\to\infty)= \psi_0 r^{-\Delta}$. For $\Delta\equiv \Delta_+ \ge 2$
we have $\psi_-=0$, $\psi_+\equiv \psi_0\neq 0$, while
for $\Delta\equiv \Delta_- < 2$ we let $\psi_+=0$, $\psi_-\equiv\psi_0\neq 0$.
The case $\Delta\equiv \Delta_{\pm}$ corresponds
to the saturation of the $5$-dimensional BF bound. 
Next to these conditions we have to impose the regularity condition
(\ref{bc_horizon}). Since the equations are invariant under the rescaling
$\psi(r)\rightarrow \psi(r)/\sqrt{\gamma}$
the normalisation of $\psi(r)$ has to be fixed in addition. In the following we
will choose $\psi(r_+)=1$.
These are a total of {\it three} boundary conditions to be fulfilled for the
{\it second} order differential equation (\ref{eq3}). 
Regular solutions corresponding to a fixed value of $\Delta$ will hence only
exist for specific values of 
$r_+/L_{\rm eff}\equiv 1/L_{\rm eff}$. 
In the following, we will let our numerical routine determine $L^2_{\rm eff}$
for a fixed value of $\Delta$
subject to the three boundary conditions mentioned above.
Clearly, the parameters $L^2$ and $m^2$ can then be reconstructed from the
numerical value of $L^2_{\rm eff}$.

\subsection{Linear case: $\gamma = 0$}
We have first studied the case $\gamma=0$, in which (\ref{eq3}) becomes linear
and
describes a scalar field in the background of the black hole solution given by
(\ref{solu}) with $a(r)\equiv 1$.

\begin{figure}
\centering
\epsfysize=8cm
\mbox{\epsffile{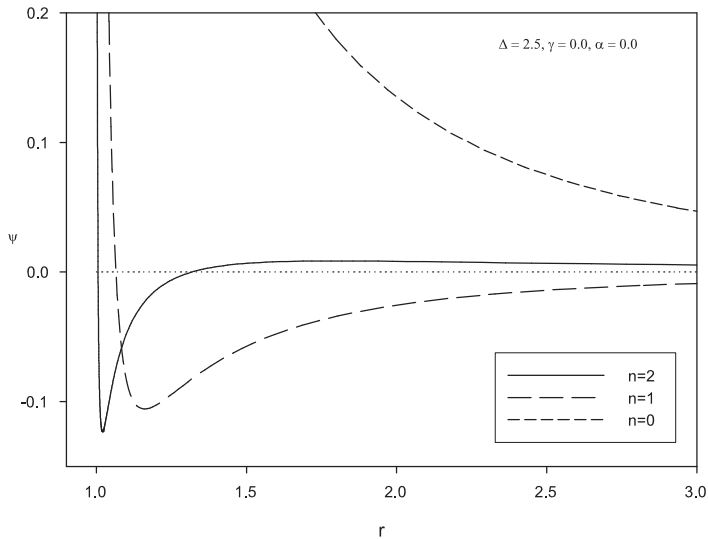}}
\caption{\label{nodes}
The profile of the scalar field function $\psi(r)$ is shown for $\Delta=2.5$,
$\gamma=0$ and $\alpha=0$. 
Next to the fundamental solution without nodes ($n=0$) there exist solutions
with nodes. We here
present the solution with $n=1$ nodes and that with $n=2$ nodes, respectively.} 
\end{figure}

\begin{figure}
\centering
\epsfysize=8cm
\mbox{\epsffile{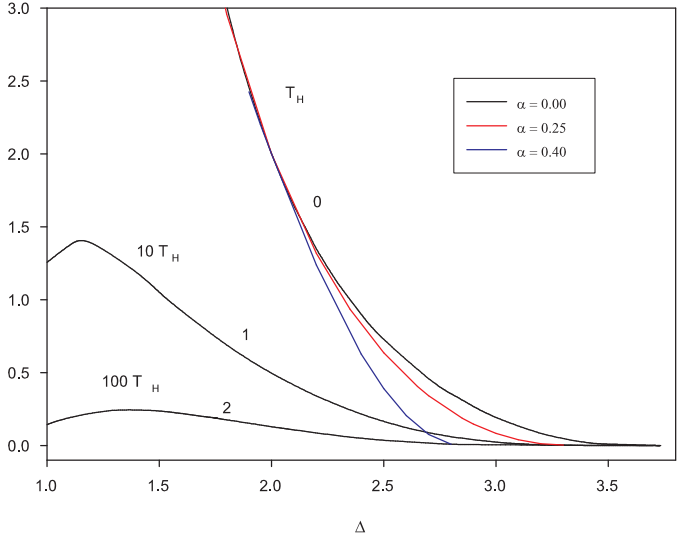}}
\caption{\label{delta_th}
The Hawking temperature $T_{\rm H}$ as function of $\Delta$ for the $n=0,1,2$
SAdS solutions with $\alpha=0$ (black).
The number of nodes $n$ is indexed by the numbers $0,1,2$, respectively, on the
curves.
We also give the temperature of the $n=0$ GBAdS black holes for $\alpha=0.25$
(red) and $\alpha=0.4$ (blue). Here $\gamma=0$.} 
\end{figure}

\begin{figure}
\centering
\epsfysize=8cm
\mbox{\epsffile{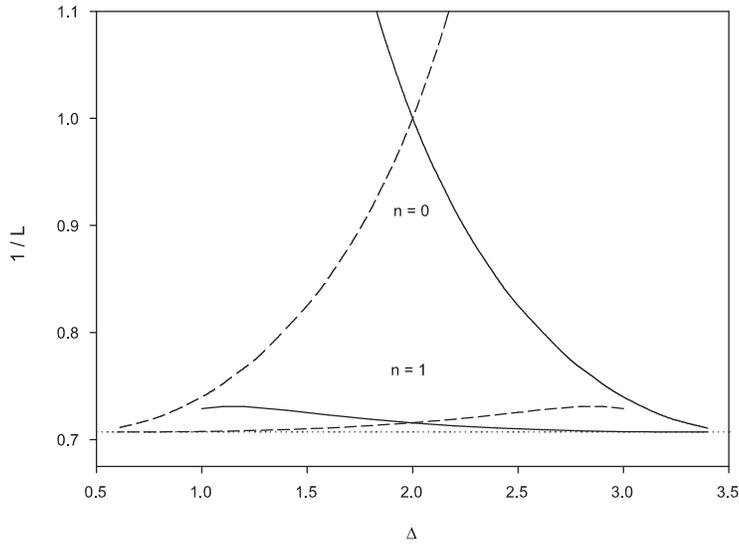}}
\caption{\label{delta}
The value of $\frac{r_+}{L}\equiv \frac{1}{L}$ as function of $\Delta$  
for the $n=0,1$ SAdS black holes with $\gamma=0$. We show the value of
$\Delta\equiv \Delta_+$ given by the solid and dashed curves
with $\Delta > 2$ and the value of $\Delta \equiv \Delta_-$ given by the solid
and dashed curves with
$\Delta < 2$. Here $\gamma=0$. } 
\end{figure}

\begin{figure}
\centering
\epsfysize=8cm
\mbox{\epsffile{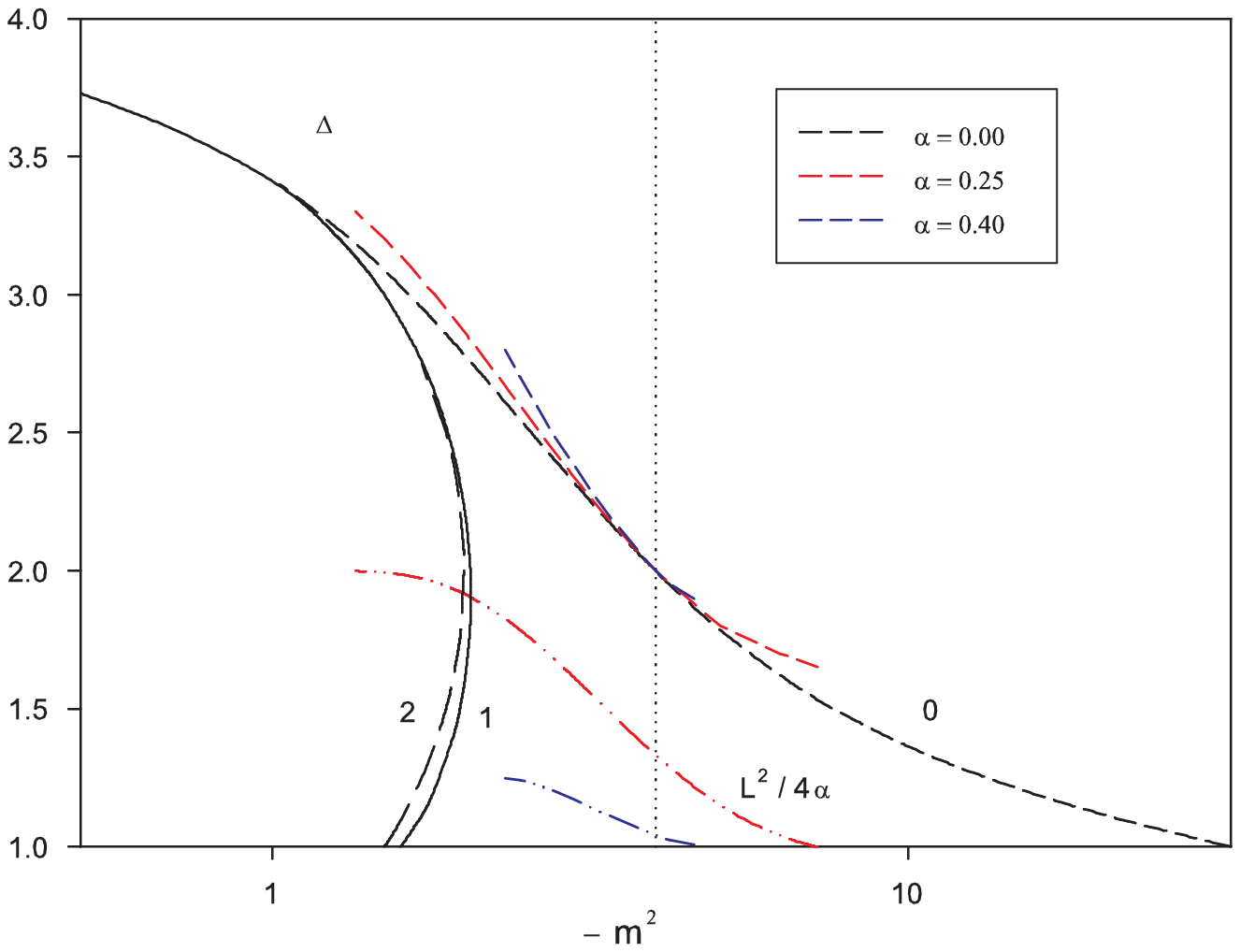}}
\caption{\label{mu2_delta}
The value of $\Delta$ as function of the mass $-m^2$ for the $n=0,1,2$
(short-dashed, solid, long-dashed) SAdS black hole solutions (black).
We also show the value of $\Delta$ for $n=0$ GBAdS black hole solutions for
$\alpha=0.25$ (red short-dashed) and
$\alpha=0.4$ (blue short-dashed), respectively. The value of $L^2/4\alpha$ is
also given for $n=0$ GB black holes
for $\alpha=0.25$ (red dotted-dashed) and $\alpha=0.4$ (blue dotted-dashed),
respectively. The dotted vertical line corresponds
to the saturation of the AdS$_5$ BF bound with $m^2=-4/L^2\equiv -4$ and
$\Delta=2$ (see also Fig. \ref{delta}). Here $\gamma=0$.} 
\end{figure}

To test our numerical technique we have reexamined the case $\alpha=0$ and
recovered the results given in \cite{Dias:2010ma}.
In addition to the hairy black hole solution constructed in \cite{Dias:2010ma},
where the scalar field $\psi(r)$ is monotonically
decreasing from $r=r_+$ to $r=\infty$, we have been able to construct solutions
where the scalar field possesses a number
$n$ of nodes. This is shown in Fig.\ref{nodes}, where we present $\psi(r)$ for
$\Delta=2.5$, $n=0,1,2$, respectively.
We believe that solutions with an arbitrary number of nodes $n$ exist, but we
did not attempt to construct the solutions
for $n > 2$.

In Fig. \ref{delta_th} we give the temperature $T_{\rm H}$ of the hairy black
holes for $n=0,1,2$ 
as function of $\Delta$. Our numerical
results suggest that the solutions stop existing for $\Delta = 2 +
\sqrt{3}\approx 3.732$.
This agrees  with the exact result when inserting the AdS$_2$ BF bound
$m^2=-1/L^2$ for $\alpha=0$ (see (\ref{bf})) into
the expression for $\Delta$ (see (\ref{lambda})). This has already been observed
in \cite{Dias:2010ma} for $n=0$, but we find
that this remains valid for solutions with nodes. Fig. \ref{delta_th} clearly
shows that the condensation
of scalar fields with nodes happens at much lower temperatures than that of the
scalar field without nodes.
In fact, we find that the higher $n$ the lower the temperature has to be and
hence the closer to the extremal solution
we have to be to find the scalar field condensation instability.

In Fig.\ref{delta} we give the value of $r_+/L\equiv 1/L$ as function of
$\Delta$ for the $n=0,1$ node solutions.
The solid $n=0$ line was already presented in \cite{Dias:2010ma}. Note that for
$\Delta > 2$, this curve
corresponds to $\Delta_+$, while for $\Delta < 2$ this is $\Delta_-$. In
addition to what was given in \cite{Dias:2010ma} we
also give the results for the other normalisable mode which has $\Delta_-$ for
$\Delta < 2$ and $\Delta_+$ for $\Delta >2$, 
respectively. We observe that in analogy to what has been observed for the $n=0$
black holes \cite{Dias:2010ma}
also $n=1$ black holes show an instability against scalar hair condensation only
for 
$r_+/L \ge r_+^{\rm(ex)}/L=\sqrt{2}$. This happens at the saturation of the
AdS$_2$
BF bound with $\Delta\equiv \Delta_+ = 2+\sqrt{3}$ or $\Delta\equiv
\Delta_-=2-\sqrt{3}$. Note, however, that
this latter value would correspond to a non-normalisable mode since $\Delta <
1$. 
The $\Delta_+$ and $\Delta_-$ curves meet at $\Delta=2$. This corresponds to the
saturation of the $5$-dimensional
BF bound with $m^2=-4/L^2$ and for the $n=0$ solution happens at $r_+=L$, i.e.
$M=0$. In \cite{Dias:2010ma} this was related
to the existence of an exact solution to the equations of motion with
$\psi(r)=L^2/r^2$ and $f(r)=-1+r^2/L^2$. 
This describes a scalar field in a space-time isometric to a part of AdS$_5$. 
For the $n=1$ solutions we find that the $\Delta_+$ and $\Delta_-$ curves meet
at $\Delta=2$, i.e. in the
limit of saturation of the AdS$_5$ BF bound, but for $r_+/L\approx 0.715$.
Hence the space-time is clearly not isometric to AdS$_5$ in this case and we
could not find any exact solution
given in terms of powers of $r$. 

Note that all black holes with $r_+/L < 1$ have $M < 0$. We observe, however,
that 
in addition to what was discussed in \cite{Dias:2010ma} $n=0$ black holes with
$r_+/L > 1$ can 
also have a scalar field condensation instability with $\Delta > 2$.
These black holes have positive $M$ and hence positive energy $E$. 
We thus find that a scalar condensation instability with $\Delta > 2$ does not
only occur on SAdS black holes with
negative energy $E$, but also on SAdS black holes with positive energy $E$. 
The $n=1$ black hole solutions exist only up to a maximal value of $r_+/L$. We
find numerically that $r_+/L \lesssim 0.731$
and that the maximal value is attained at $\Delta=1.2$. Hence, scalar fields
with nodes can only condensate on small black holes.

In Fig.\ref{mu2_delta} we give $\Delta$ as function of $m^2$ for the $n=0,1,2$
solutions.
We note that while for the $n=0$ solutions the mass range 
for which scalar condensation appears is quite large,
this is different for the $n=1$ and $n=2$ node solutions. 
For all solutions, the maximal value of $m^2=m^2_{\rm max}$ corresponds to
the saturation of the $2$-dimensional BF bound $m^2=m_{\rm BF}^2=-1/L^2$ at the
extremal solution with 
$r_+=r_+^{\rm(ex)}=L/\sqrt{2}$ (see discussion above).
With our choice $r_+=1$ this corresponds to $m^2_{\rm max}=-0.5$.
The minimal value of $m^2=m^2_{\rm min}$ on the other hand depends to the number
of nodes $n$.
With the restriction
that $\Delta \ge 1$ we find $m^2_{\rm min}\approx -32.1$ with corresponding
$r_+/L\equiv 1/L\approx 3.20$  for $n=0$,
$m^2_{\rm min}\approx -2.48$ with corresponding $r_+/L\equiv 1/L\approx 0.72$
for $n=1$ and 
$m^2_{\rm min}\approx -2.01$ with corresponding $r_+/L\equiv 1/L\approx 0.71$
for $n=2$. Note that the minimal
value of $m^2$ is attained at $\Delta=1$ for the $n=0$ solutions, but at
$\Delta\approx 2$ for the $n=1,2$ solutions.
The value of $m^2$ at $\Delta=1$ is larger than $m^2_{\rm min}$ for the
solutions with nodes and is
$m^2_{\Delta=1}\approx -1.59$ for the $n=1$ solution and $m^2_{\Delta=1}\approx
-1.50$  for the $n=2$ 
solution, respectively.
We thus conclude that for a fixed value of the mass in the interval $m^2_{\rm
min} \le m^2 \le m^2_{\Delta=1}$ there exist
two hairy black hole solutions which differ in their fall-off of the scalar
field function on the AdS boundary.
Moreover, the interval of the mass $m^2$ on which scalar condensation appears
decreases with the increase
of the number of nodes $n$. 

Next, we have studied the influence of the Gauss-Bonnet term by letting
$\alpha\neq 0$. We find that the solutions
with nodes also exist in this case. However, the numerical values for the
interesting quantities 
are very close to those for $\alpha=0$. This is why we do not present them
here. 
We give the temperature $T_{\rm H}$ of the $n=0$ GBAdS solutions
for $\alpha=0.25$ and $\alpha=0.4$ in Fig.\ref{delta_th}. We observe that the
solutions cease to exist for
smaller values of $\Delta$ as compared to the $\alpha=0$ limit. This can be
explained by considering (\ref{bf}) 
for $\alpha\neq 0$ and inserting $m^2=-1/(L^2-4\alpha)$ into
the expression for $\Delta$. This is a complicated expression, but the effect
can be seen easily when expanding $\Delta$ for small
$\alpha$. This reads $\Delta\approx 2+ \sqrt{3-3\alpha/L^2 -11\alpha^2/L^4 +
O(\alpha^3)}$ and clearly
shows that $\Delta$ decreases with increasing $\alpha$. Moreover we find perfect
agreement between
the exact value of $\Delta$ and our numerical results: $\Delta\approx 3.51$ for
$\alpha=0.25$ and $\Delta\approx 2.62$ for $\alpha=0.4$, respectively.

In Fig.\ref{mu2_delta} we give $\Delta$ and the value of $L^2/4$ as function of
$m^2$ for the $n=0$ GBAdS black holes
for $\alpha=0.25$ and $\alpha=0.4$, respectively. 
For $\alpha\neq 0$, we find that the maximal value of $m^2=m^2_{\rm max}$ is
again connected to the saturation 
of the AdS$_2$
BF bound $m^2=-1/(L^2-4\alpha)$ in the extremal limit with
$r_+=r_+^{\rm(ex)}=L/\sqrt{2}$. With our choice $r_+=1$ which
corresponds to $L=\sqrt{2}$ for the extremal solution we find $m^2_{\rm max} =
-1$ for $\alpha=0.25$ and $m^2_{\rm max}= 
-2.5$ for $\alpha=0.4$. This is in good agreement with the results
shown in Fig.\ref{mu2_delta} and can also be seen from the curves for
$L^2/(4\alpha)$. In the extremal
limit we have $L=\sqrt{2}$ with our choice $r_+=1$. Hence, $L^2/(4\alpha)=2$ for
$\alpha=0.25$ and $L^2/(4\alpha)=1.25$
for $\alpha=0.4$ for the largest possible value of $m^2$. This is in perfect
agreement with our numerical
results. We also find that there is a minimal value of $m^2=m^2_{\rm min}$ up to
where scalar
condensation appears. This can be understood when looking at the curves
$L^2/(4\alpha)$. These end
at $L^2/(4\alpha)=1$, which corresponds to the Chern-Simons limit. For
$L^2/(4\alpha) > 1$ there exists no
black hole interpretation of the solutions, hence it is natural that these
curves end here.
For the minimal value of $m^2=m^2_{\rm min}$ we find that $m^2_{\rm min} \approx
-7.2$ 
for $\alpha=0.25$
and $m^2_{\rm min} \approx -4.6$ for $\alpha=0.4$.
We hence observe that the interval of $m^2$ on which the scalar condensation
instability appears decreases
considerably when including the GB term.


\subsection{Non-linear case: $\gamma  \neq  0$}

\begin{figure}
\centering
\epsfysize=8cm
\mbox{\epsffile{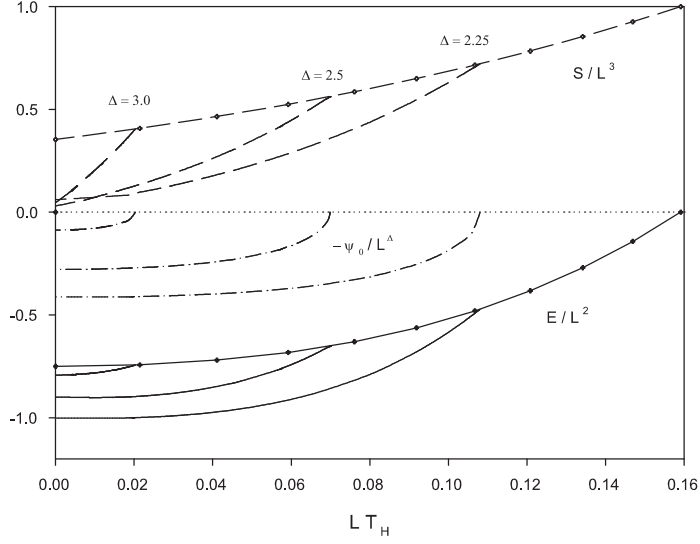}}
\caption{\label{thermo1}
The dimensionless entropy $S/L^3$ (dashed), the dimensionless energy $E/L^2$
(solid)  and the value of 
$\psi_0/L^{\Delta}$ (dotted-dashed) are given
as function of the dimensionless temperature $L T_{\rm H}$ for different values
of $\Delta$ and $\alpha=0$. 
The dashed and solid lines with additional small black dots correspond to the
entropy and energy
of the exact solution (\ref{solu}) which has $\psi(r)\equiv 0$. Here
$\gamma=1$.} 
\end{figure}

\begin{figure}
\centering
\epsfysize=8cm
\mbox{\epsffile{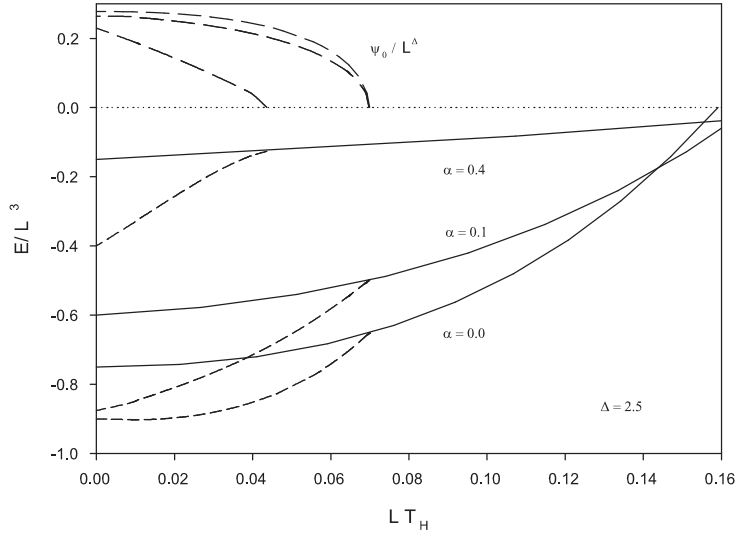}}
\caption{\label{thermo2a}
The dimensionless energy $E/L^2$ of the exact solution (\ref{solu}) (solid) as
well as of the black hole
solutions with scalar hair (short-dashed) is given
as function of the dimensionless temperature $L T_{\rm H}$ for $\Delta=2.5$ and
different values of $\alpha$. Also shown
is the value of $\psi_0/L^{\Delta}$ (long-dashed) which becomes non-equal to
zero at the value of $LT_{\rm H}$ at which
the short-dashed curves branch of from the solid curves. This corresponds to the
onset of the scalar condensation instability.
Here $\gamma=1$.} 
\end{figure}

\begin{figure}
\centering
\epsfysize=8cm
\mbox{\epsffile{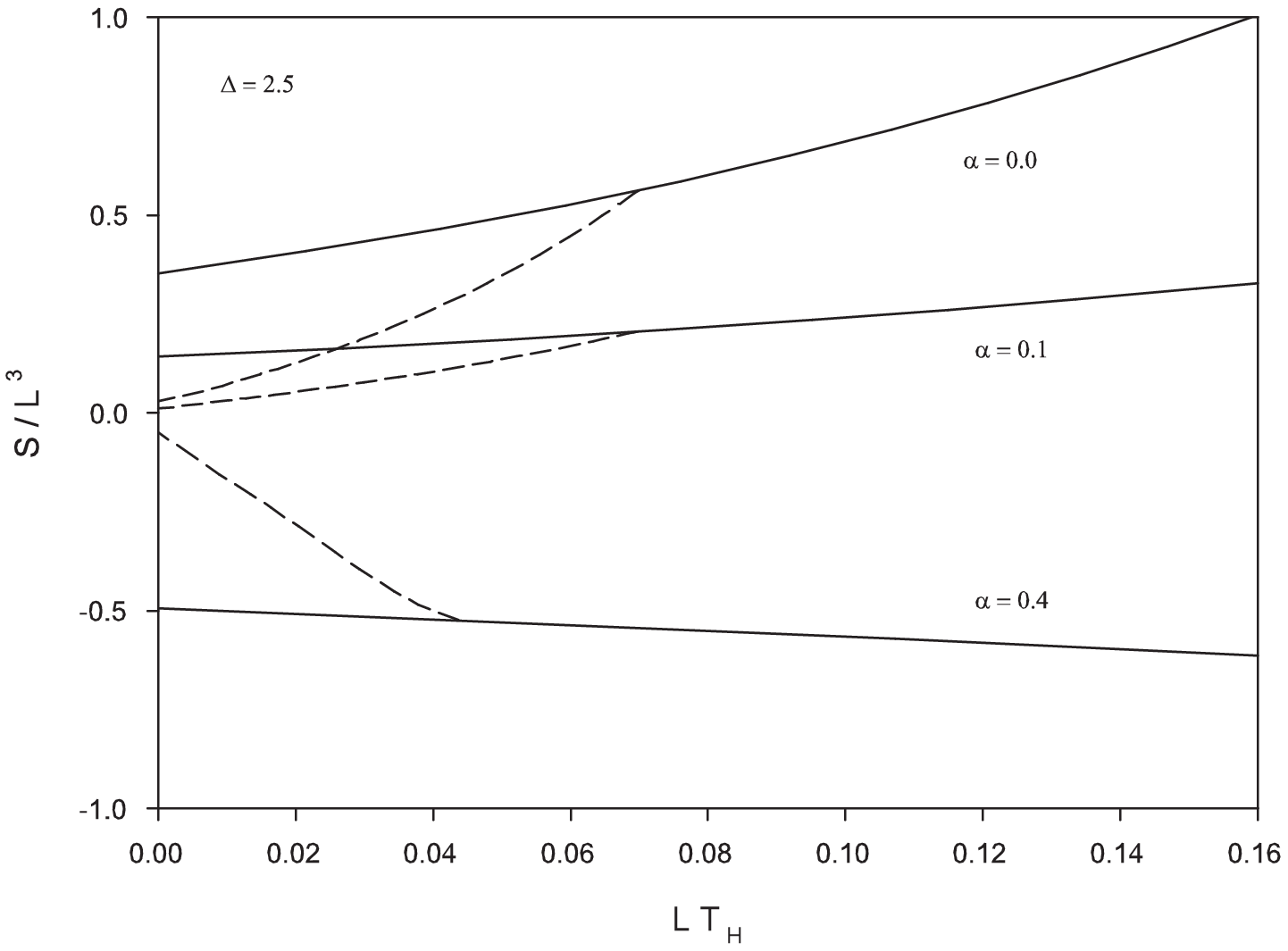}}
\caption{\label{thermo2b}
The dimensionless entropy $S/L^3$ of the exact solution (\ref{solu}) (solid) as
well as of the black hole
solutions with scalar hair (short-dashed) are given
as function of the dimensionless temperature $L T_{\rm H}$ for $\Delta=2.5$ and
different values of $\alpha$. The short-dashed
curves branch of from the solid curves at the onset of the scalar condensation
instability. Here $\gamma=1$.} 
\end{figure}

For $\gamma\neq 0$, we have to solve the full set of equations
(\ref{eq1})-(\ref{eq3}) subject to the
appropriate boundary conditions. The scalar field backreacts onto the space-time
in this case.
As for $\gamma=0$ we also find that solutions with $n$ nodes of the scalar
field function exist. However, we will only present our results for the $n=0$
solutions in this section.

As mentioned earlier we can absorb $\gamma$ into the normalisation of $\psi(r)$
by rescaling $\psi(r)\rightarrow \psi(r)/\sqrt{\gamma}$. 
We can hence set $\gamma=1$ without loss of generality and label the solution by
the value of
$\psi_0$.  Solving the equations for fixed values of $\Delta$ and different
values of $\psi_0$ we find a family
of black hole solutions with a definite relation between $\psi_0$ 
and the AdS radius $L$.  Again, we have first reinvestigated the $\alpha=0$ case
and
our results are shown in Fig.\ref{thermo1}, where we give the dimensionless
entropy $S/L^3$, the dimensionless
energy $E/L^2$ and the value of $\psi_0/L^{\Delta}$ as function of the
dimensionless Hawking temperature $LT_{\rm H}$ for different
values of $\Delta$. The case $\Delta =2$ has been given in \cite{Dias:2010ma}.
Here we present the black hole quantities
for $\Delta=2.25$, $\Delta=2.5$ and $\Delta=3$, respectively. We find that at
the onset of the scalar condensation
instability which corresponds to the value of $L T_{\rm H}=(LT_{\rm H})_{\rm
cr}$ at which $\psi_0$ becomes non-vanishing the curves
for the entropy $S/L^3$ and entropy $E/L^2$ branch of from the curves of the exact
solution given by (\ref{solu}), $a(r)\equiv 1$ 
and $\psi(r)\equiv 0$. We find that the higher the value of $\Delta$ the lower
$(LT_{\rm H})_{\rm cr}$:
$(LT_{\rm H})_{\rm cr}\approx 0.11$ for $\Delta=2.25$, $(LT_{\rm H})_{\rm
cr}\approx 0.07$ for $\Delta=2.5$  and 
$(LT_{\rm H})_{\rm cr}\approx 0.02$ for $\Delta=3$. To state
it differently: the stronger we want the scalar field to fall off on the AdS
boundary, the closer
we have to choose the solution to be to the extremal solution with $T_{\rm
H}=0$. All solutions with scalar
hair have entropy $S/L^3$ and energy $E/L^2$ smaller than the solutions without
hair which suggests that the
solutions without scalar hair are thermodynamically favoured. In the limit
$T_{\rm H}=0$ we find in agreement with \cite{Dias:2010ma}
that $S/L^3=0$. Hence, the hairy black hole should not be interpreted as extreme
black hole solutions in this limit \cite{Dias:2010ma}.

In Fig.\ref{thermo2a} and Fig.\ref{thermo2b} we give the dimensionless entropy
$S/L^3$, the dimensionless
energy $E/L^2$ and the value of $\psi_0/L^{\Delta}$ as function of the
dimensionless Hawking temperature $LT_{\rm H}$ for different
values of $\alpha$ and $\Delta=2.5$. Again, the curves for the energy $E/L^2$
and the entropy $S/L^3$ branch of from
those for the exact solution (\ref{solu}) at the onset of the scalar
condensation instability at which $\psi_0$ becomes
non-vanishing. Note that we use the ``standard`` definition for $S$ making the
entropy negative
for certain GBAdS black holes. We could substract the entropy of the $M=0$
solution from this as 
was suggested in \cite{Clunan:2004tb}, but this would not change the qualitative
results and moreover is still an open question.
Our numerical results suggest that the larger $\alpha$ the smaller $(LT_{\rm
H})_{\rm cr}$ at which scalar condensation
appears: $(LT_{\rm H})_{\rm cr}\approx 0.07$ for $\alpha=0$, $(LT_{\rm H})_{\rm
cr}\approx 0.065$ for $\alpha=0.25$
and $(LT_{\rm H})_{\rm cr}\approx 0.029$ for $\alpha=0.4$. Stating it
differently: the larger $\alpha$ the harder it gets to make an uncharged scalar
field 
condense on a black hole. This has already been observed in the case of charged
scalar fields in the context of
holographic superconductors \cite{Gregory:2009fj,Brihaye:2010mr}.

\section{Conclusions}
\label{conclusions}
In this paper we have studied the condensation of an uncharged scalar field on
uncharged hyperbolic SAdS and GBAdS black holes
in $(4+1)$ dimensions. These black holes possess an extremal limit such that the
near horizon geometry contains an 
AdS$_2$ factor. We confirm that the onset of scalar condensation appears for
scalar field mass equal to the
AdS$_2$ BF bound at the extremal solution. Near extremal black holes can also
form a scalar condensate, however, the
fall-off of the scalar field on the AdS boundary is the slower the further we
move away from extremality. 
We find that a family of hairy black hole solutions exists which is labelled by
the number of nodes of the
scalar field function. We observe that the higher the number of nodes the closer
we have to be to extremality to
find scalar field condensation. Moreover, scalar fields with nodes can only
condense
on reasonably small black holes. We also observe that the inclusion of the GB
term makes scalar field condensation harder and restricts
the range of masses of the scalar field for which hairy black holes exist
considerably.  

In \cite{Dias:2010ma} static, hyperbolic SAdS black holes were studied as toy
model for rotating black holes. It was
shown in \cite{Reall:2002bh} that any extremal black hole admits a near-horizon
limit. 
Hence, we would expect that for rotating, near-extremal GB black holes in
asymptotically flat space
similar instabilities against scalar condensation would appear and this is
currently under investigation. 
Moreover, it would be interesting to see whether there exist also hairy black
holes with charged
scalar fields that contain nodes. Within the context of holographic
superconductors these different black holes
could be the gravity duals of different superconducting phases at low enough
temperature. 
\\
\\
{\bf Acknowledgments}
YB would like to thank the Belgian FNRS for financial support.


\appendix
\section{Appendix: Extremal solutions}
Let us assume that an extremal limit for black hole solutions in AdS$_5$ exists
such that the near horizon
geometry is given by AdS$_2\times M_{3}$, where $M_3$ can be a manifold of
positive curvature
($k=1$), negative curvature ($k=-1$) or zero curvature ($k=0$). Let the volume
of $M_3$ be denoted by $V_3$.
Note that only hyperbolic GBAdS and SAdS black holes ($k=-1$) possess an
extremal
limit. For the cases $k=1$ and $k=0$, however, one could add an electric charge
to obtain an extremal limit.
The entropy function formalism \cite{Sen:2005wa,sen2,dias_silva} was used for
charged GBAdS black holes with
spherical horizon ($k=1$) in \cite{aste}. Here, we will extend these results to
the cases $k=0$ and $k=-1$ keeping
in mind that we are interested in the uncharged limit with $k=-1$ in this paper.
Following \cite{aste} we write the metric of the AdS$_2\times M_3$ space-time as
\begin{equation}
 ds^2=v_1 \left(-\rho^2 d\tau^2 + \frac{1}{\rho^2} d\rho^2\right) + v_2
d\Sigma^2_{k,3}  \ ,
\end{equation}
where 
\begin{equation}
d\Sigma^2_{k,3}=
\begin{cases}
d\Omega^2_{3}=d\psi^2 + \sin^2\psi\left(d\theta^2 + \sin^2\theta
d\varphi^2\right) \ \ \ \ \ {\rm for} \ \ k=1 \\ 
dx^2 + dy^2 + dz^2 \hspace{3.85cm} {\rm for} \ \
k=0 \\ 
d\Xi_3^2=d\psi^2 + \sinh^2\psi \left(d\theta^2 + \sin^2\theta d\varphi^2\right)\
\ \  {\rm for} \ \  k=-1   \    
\end{cases}
\end{equation}
The entropy function is $F(v_1,v_2,e,Q) \propto Qe-f(v_1,v_2,e)$, where
$f(v_1,v_2,e)$ corresponds to the integral of the action density over the
coordinates of $M_3$:
\begin{equation}
 f(v_1,v_2,e)=\int\limits_{M_3} \sqrt{-g} \tilde{{\cal L}} d^3 x \ .
\end{equation}
$\tilde{{\cal L}}$ is the Lagrangian density of Einstein-Gauss-Bonnet-Maxwell
theory in AdS
\begin{equation}
 \tilde{{\cal L}}=\left(R -2\Lambda + 
\frac{\alpha}{2}\left(R^{MNKL} R_{MNKL} - 4 R^{MN}
R_{MN} + R^2\right) -F_{MN} F^{MN} \right) \ , \ M,N,K,L=0,1,2,3,4 \ .
\end{equation}
$F_{MN}$ is the field strength tensor of a U(1) gauge field, which we will
choose as $F_{\tau\rho}=-F_{\rho\tau}=e$ \cite{aste}. Inserting
the metric and the field strength tensor we find
\begin{equation}
 f(v_1,v_2,e)=V_3 \left(6kv_1 \sqrt{v_2} -2 v_2^{3/2} + v_1 v_2^{3/2}
\frac{12}{L^2} - 
12k \alpha \sqrt{v_2} + 2e^2 \frac{v_2^{3/2}}{v_1}\right) \ , 
\end{equation}
where $V_3$ corresponds to the integral over the coordinates of $M_3$, which for
$k=1$ is $V_3=2\pi^2$, while
we set $V_3=1$ for $k=0$ and $k=-1$ such that all quantities that contain this
factor are given per unit
volume.
  
The attractor equations are
\begin{eqnarray}
 \frac{\partial F}{\partial v_1}=0 &\rightarrow& 6kv_1^2 + \frac{12}{L^2} v_1^2
v_2 - 2 e^2 v_2 =0 \ , \\
\frac{\partial F}{\partial v_2}=0 &\rightarrow&  kv_1^2 - v_1 v_2 +
\frac{6}{L^2} v_1^2 v_2 - 2k \alpha v_1 + e^2 v_2 =0 \ , \\
\frac{\partial F}{\partial e}=0 &\rightarrow& Q=4 V_3 e \frac{v_2^{3/2}}{v_1} \
,
\end{eqnarray}
which reduces to the equations given in \cite{aste} for $k=1$. 
With the identifications $v_2=r_+^2$, $v_1=R^2$, where $R$ is the radius of the
AdS$_2$ we then find
\begin{equation}
\label{charge}
 \frac{Q^2}{16 V_3^2}= 3 r_+^4 \left(k + \frac{2 r_+^2}{L^2}\right) 
\end{equation}
and 
\begin{equation}
\label{ads2radius}
 R^2 = \frac{r_+^2 + 2 k \alpha}{4k + 12r_+^2/L^2} \ . 
\end{equation}
Note that $Q$ is strictly positive for $k=0$, $k=1$, which relates to the
statement above that extremal
black holes with flat or spherical horizons exist only if charge is included. On
the other hand, for hyperbolic
horizons, i.e. $k=-1$ we can set $Q=0$ and find that the value of the horizon is
at $r_+=L/\sqrt{2}$. We hence
find the following values for an extremal GBAdS black hole with hyperbolic
horizon and vanishing charge
that possesses a near horizon geometry of AdS$_2\times M_3\equiv$ AdS$_2\times
H^3$~:
\begin{equation}
 Q=0 \ \ , \ \ r_+=L/\sqrt{2} \ \ , \ \ R^2 = \frac{L^2}{4} - \alpha \ .
\end{equation}
For $\alpha=0$ this reduces to the well-known SAdS limit with $R=L/2$.

\end{document}